\documentclass[12pt]{iopart}
\usepackage{iopams}  

\expandafter\let\csname equation*\endcsname\relax
\expandafter\let\csname endequation*\endcsname\relax

\usepackage{amsmath}
\usepackage{amssymb} 
\usepackage[cm]{fullpage}
\usepackage{graphicx} 
\usepackage{siunitx}
\usepackage{tabularx} 
\usepackage[numbers, super]{natbib}
\usepackage[colorlinks,linkcolor=blue,citecolor=blue,urlcolor=blue]{hyperref}
\usepackage[table,xcdraw]{xcolor}
\usepackage[utf8]{inputenc}

\DeclareSIUnit\parsec{pc}

\newcommand{\beq}{\begin{equation}}
\newcommand{\eeq}{\end{equation}\\}
\newcommand{\beqa}{\begin{eqnarray}}
\newcommand{\eeqa}{\end{eqnarray}\\}

\newcommand{\rpar}[1]{\left(#1\right)}
\newcommand{\spar}[1]{\left[#1\right]}

\newcommand{\bbox}[1]{\rpar{\Box #1}}
\newcommand{\bd}{{\mathrm d}}

\newcommand{\HLK}[1]{#1}
\newcommand{\HLV}[1]{#1}
\newcommand{\HLs}[1]{#1}
\newcommand{\HLy}[1]{#1}
\newcommand{\HLyx}[1]{#1}
\newcommand{\HLg}[1]{#1}

\begin{document}

\title{Einstein-Gauss-Bonnet gravity with extra dimensions}

\author{Carsten van de Bruck and Chris Longden}

\address{Consortium  for  Fundamental  Physics,  School  of  Mathematics  and  Statistics,  University  of  Sheffield,
Hounsfield Road, Sheffield, S3 7RH, UK}
\ead{c.vandebruck@sheffield.ac.uk, cjlongden1@sheffield.ac.uk}
\vspace{10pt}
\begin{indented}
\item[]September 2018
\end{indented}

\begin{abstract}
We consider a theory of modified gravity possessing $d$ extra spatial dimensions with a maximally symmetric metric and a scale factor, whose $(4+d)$-dimensional gravitational action contains terms proportional to quadratic curvature scalars. Constructing the 4D effective field theory by dimensional reduction, we find that a special case of our action where the additional terms appear in the well-known Gauss-Bonnet combination is of special interest as it uniquely produces a Horndeski scalar-tensor theory in the 4D effective action. We further consider the possibility of achieving stabilised extra dimensions in this scenario, as a function of the number and curvature of extra dimensions, as well as the strength of the Gauss-Bonnet coupling. Further questions that remain to be answered such as the influence of matter-coupling are briefly discussed.
\end{abstract}

%\submitto{\CQG}
\maketitle

\section{Introduction}

There is no concrete \emph{a priori} reason why the spacetime we live in should have precisely three spatial dimensions and one time dimension (for overviews of diverse theories with extra dimensions and their physical consequences see e.g. \cite{Zumino:1985dp, Overduin:1998pn, Rubakov:2001kp, Brax:2003fv}. Instead, theoretical physicists hope that a fundamental theory of nature will be able to predict the number of dimensions of spacetime. In string theory, for example, consistency requires additional spatial dimensions. At the very least, a theory proposing more than three spatial dimensions requires the additional dimensions to be either small in order to be physically plausible or require objects like branes, on which standard model particles are confined, in order for the theory to be consistent with observations. A central question in such theories is, then, how these extra dimensions become sufficiently small and/or stabilise, given the field content and dynamics of such theories.  

It has been established for a long time that extra-dimensional theories can, in the appropriate limit, behave like a conventional four-dimensional spacetime with additional field content derived from the influence of the extra dimensions. Kaluza-Klein theory, in which one obtains gravity and electromagnetism in four dimensions by starting from gravity alone in five dimensions, is the textbook example. More modern realisations of this paradigm, including the work to be presented in what follows, often parametrise the size of the extra dimensions with a scalar field, leading to an effective 4D action which is a scalar-tensor theory. Similarly, one can consider, particularly in string theory contexts, the influence of D-branes and other objects found in the extra-dimensional space on our perceived four-dimensional world to produce scalar-tensor and vector-tensor theories with interesting properties.

Extra dimensions, as well as being of general interest in theoretical physics, have particular applications and consequences in the context of cosmology. Questions on the nature of dark matter and dark energy could potentially have resolutions in theories of matter existing in hidden extra dimensions, or of extra-dimensional effects on the dynamics of the observed four-dimensional universe. Similarly, early-universe phenomena such as inflation could be driven by extra-dimensional effects, or at least occur in the presence of extra-dimensions. For work in these directions, see \cite{Kanti:2015pda, Wongjun:2013jna, Nojiri:2005vv, Amendola:2005cr, Amendola:2007ni, Koivisto:2006ai, Neupane:2006dp,Leith:2007bu,Odintsov:2018zhw}. 

For many purposes of interest, it is useful if extra dimensions can be stabilised in size, at least to high precision. Not only does this come with the desirable property of avoiding problems associated with an endless collapse down to the uncharted territory of poorly understood Planck-scale effects or the evolution of fundamental constants, but it also lends itself well to, say, application to dark energy. Cosmological observations indicate that that the energy density of dark energy is either very close to, or exactly, constant, and in scalar-tensor theories, this is often due to a frozen, non-evolving, scalar field. When thinking about extra dimensions, this is most immediately understood, then, as a constant-size extra-dimensional space. A constant or near-constant energy density is also desirable from the perspective of inflation.

As the introduction of additional spatial dimensions often modifies gravity in the four-dimensional effective action, there will unavoidably be questions regarding the compatibility of such theories with local constraints on deviations from General Relativity (GR). This is another problem that may be alleviated if the extra dimensions are sufficiently stabilised that they do little more than contribute vacuum energy at present, while keeping open the possibility that larger and more dynamic extra-dimensional effects were present in the early universe where constraints are weaker.

Unfortunately, it is not a trivial task to stabilise extra dimensions. If gravity in the full extra-dimensional theory is described by GR one typically does not find such situations in which one (or several) dimensions are stabilised. In fact, it has been argued that it is highly non--trivial in GR to stabilise extra dimensions \cite{Carroll:2001ih}. Instead, within GR, all dimensions want to be dynamical (for a disucssion on Kaluza--Klein gravity and its consequences, see \cite{Overduin:1998pn}). This motivates our work in this paper in which we study an extended theory of gravity. In particular, we add terms proportional to all quadratic curvature scalars to the action. From an effective field theory perspective, it is these terms that we would expect to bring leading order corrections to GR, and they are indeed often generated in high energy theories (for a discussion see e.g. \cite{Garraffo:2008hu}). Starobinsky's $R^2$ gravity is the most well-known example. It is the hope that these contributions generate an effective potential for the scale factor of the extra dimensions, which in the low-energy effective field theory is a scalar field. 

A similarly ubiquitous special case known as the Gauss-Bonnet (GB) combination will be of particular importance to this work. The GB term is unique in four dimensions as a total derivative which makes no classical contribution to the dynamics when uncoupled, but in higher-dimensional spacetimes has no such property and can instead be expected to have a pronounced effect. As well as this conceptual relevance to the topic of extra dimensions, GB corrections appear in string theories and have been considered in the context of cosmology when coupled to a scalar-tensor theory. Previous work in this direction found that the GB term can impede or even freeze the motion of a scalar field \cite{vandeBruck:2015gjd, vandeBruck:2016xvt, vandeBruck:2017voa}. One question that arises from this, motivating the present work, is whether it could perhaps play a similar role in extra-dimensional scenarios and stabilise the size of the additional dimensions? Questions like this have been addressed elsewhere, including work which considers the presence of branes \cite{Alberghi:2005vq, Charmousis:2002rc, Charmousis:2003ke, Elizalde:2006ub, Canfora:2013xsa, Canfora:2014iga}. Here we wish to study the four-dimensional effective theory and calculate the effective potential generated by corrections to the Einstein-Hilbert action. We assume that the higher--dimensional spacetime is $4 + d$--dimensional, and the dynamics of the $d$--dimensional additional space is described by a single--scalar factor $b(x)$. While more complicated scenarios are possible to study, our work will shed light about the effective potential for the 'scalar field' $b$, generated in the low-energy effective action. Our motivation is not dark energy or modified gravity, but rather whether the GB induces a potential in the 4D effective theory such that the extra dimensions stabilise. The recent detection of gravitational waves and the electromagnetic counterpart of the binary neutron star merger GW170817\cite{GBM:2017lvd} implies that gravitational waves propagate with the speed of light. A GB correction which is important at late times would not be compatible with this results and therefore the extra--dimensions have to be frozen at late times in a set-up described in this paper \cite{Ezquiaga:2017ekz,Creminelli:2017sry,Sakstein:2017xjx}. 

The paper is organised as follows. In the next Section, we describe the general setup in $4+d$ dimensions and determine the four-dimensional effective action. As we will see, in the case of a $4+d$--dimensional Einstein--Gauss-Bonnet theory, the effective four-dimensional theory is of Horndeski form, in which the scale factor describing the size of the extra dimensional submanifold is a scalar field. In Section 3 we discuss some properties of the resulting 4D theory. In particular we look for minima in the effective potential. We also briefly comment on the inclusion of matter. Our conclusions can be found in Section 4. 

\section{A theory of modified gravity with extra dimensions} \label{sec:model}

We first consider a theory of gravity constructed in $4+d$ spacetime dimensions in which the usual $(4+d)$-dimensional Ricci scalar, $R$, is accompanied by an arbitrary cosmological constant and terms proportional to the three quadratic-order curvature scalars $R^2$, $R_{\mu\nu}R^{\mu\nu}$ and $R_{\rho\sigma\mu\nu}R^{\rho\sigma\mu\nu}$, that is,

\beq \label{eq:4pdaction}
S = \frac{M_{4+d}^{2+d}}{2} \int \bd^{(4+d)} X \sqrt{-G} \rpar{ R + c_1 R^2 + c_2 R_{AB} R^{AB} + c_3 R_{ABCD} R^{ABCD} - 2 \Lambda_{4+d}} \, ,
\eeq

where capital Roman indices run from $0 \ldots d+3$ and denote objects defined on the full $(4+d)$-dimensional manifold. The metric in $(4+d)$ dimensions, $G$, is to be understood as a product metric built from a 4-dimensional metric $g$ and a $d$-dimensional metric $\gamma$ such that

\beq \label{eq:4pdmetric}
\bd s^2 = G_{AB} \bd X^A \bd X^B = g_{\alpha \beta}(x) \bd x^\alpha \bd x^\beta + b(x)^2 \gamma_{mn}(y) \bd y^m \bd y^n \, ,
\eeq

where we use Greek letters and lower case Roman letters to label objects defined on the 4-dimensional (4D) and $d$-dimensional components respectively. As such, our ansatz for the $4+d$--dimensional spacetime is similar to the one used in \cite{Canfora:2013xsa}. Similarly $x^\alpha$ is used to represent the coordinates on the 4-dimensional spacetime and $y^m$ those of the $d$ extra dimensions. In the above we have introduced a scale factor $b(x)$ controlling the size of the $d$ extra dimensions which is only a function of the $4$-dimensional coordinates. For simplicity, we make the assumption that $\gamma$ is the metric of a maximally symmetric space with Ricci curvature $R_d$.

To study the 4D effective action of the above theory, we will perform dimensional reduction. In this process, one decomposes the theory into tensor and scalar parts, where the scalar field is identified with the $d$-dimensional metric's scale factor, $b$. As this only depends on $x^\alpha$ (and because of our previous assumption of maximal symmetry in $\gamma$), the part of the action depending on coordinates $y^a$ can be factored out of the action integral such that

\beq \label{eq:reducedaction}
S = \frac{M_{4+d}^{2+d}}{2} \int \bd^4 x \bd^d y \sqrt{-g} \sqrt{\gamma} b^d L_{4\text{D}} = \frac{M_{4+d}^{2+d} \mathcal{V}}{2} \int \bd^4 x \sqrt{-g} b^d L_{4\text{D}} \, ,
\eeq

where

\beq
\mathcal{V} = \int \bd^d y \sqrt{\gamma} \, ,
\eeq

is the comoving volume of the extra dimensions and $ L_{4\text{D}}$ is a four-dimensional effective Lagrangian, only depending on 4D curvature tensors and the scalar $b$. To obtain this, one uses the form of the metric $G$ in eq.(\ref{eq:4pdmetric}) to explicitly evaluate curvature tensors in terms of those of metrics $g$ and $\gamma$, as well as the scale factor $b$. After straightforward but tedious computations, one finds it can be written in the form

\beq
L_{4\text{D}} = L_4 + L_d + L_b - 2 \Lambda_{4+d} \, ,
\eeq

where $L_4$ is a four-dimensional gravity sector containing a non-minimal (kinetic) coupling between $b$ and the 4D Ricci scalar $R_4$, the same quadratic curvature terms as the original higher-dimensional action, as well as a coupling of the 4D Ricci Tensor to the derivatives of the field

\beq \label{eq:dimred-L4}
L_4 = \Bigg[ \HLy{1 + 2 c_1 \Bigg( \frac{R_d}{b^2} - \frac{d(d-1)}{b^2} (\partial b)^2}  \HLg{ \ - \ \frac{2d}{b} \bbox{b} \Bigg)} \Bigg] R_4 + c_1 R_4^2 + c_2 R_{\alpha\beta} R^{\alpha\beta}+ c_3 R_{\alpha\beta\gamma\delta} R^{\alpha\beta\gamma\delta}\HLg{\ - \  \frac{2 d c_2}{b} R_{\alpha \beta} \partial^\alpha \partial^\beta b} \, .
\eeq

Similarly, $L_d$ contains terms that look like non-minimal gravity for the $d$-dimensional part of the metric

\beq
L_d = \Bigg[ \HLV{1} \HLK{ \ - \  2 \frac{d(d-1) c_1 + (d-1) c_2 + 2 c_3}{b^2}(\partial b)^2} \HLs{ \ -  \ 2 \frac{2 d c_1 + c_2}{b} \bbox{b}} \Bigg] \frac{R_d}{b^2} \ \HLV{+ \ \spar{c_1 + \frac{c_2}{d}+ \frac{2c_3}{d(d-1)}}\frac{R_d^2}{b^4}} \, ,
\eeq

where in the last term maximal symmetry has been used to express the squares of the Ricci and Riemann tensors in terms of the constant Ricci scalar $R_d$. \footnote{This expression is invalid in the case $d=1$. The physical reason for this is that a one-dimensional space has identically zero Ricci curvature and these terms hence do not arise unless $d \geq 2$ in the first place.} Because of this, while it resembles gravity in $d$ dimensions, it is better identified as kinetic and potential terms for the scalar degree of freedom that are proportional to the extra-dimensional curvature (and its square). 

Additionally, we find terms corresponding to a Lagrangian for the $b$ field that does not depend on $R_d$, contributing further kinetic structure to the 4D theory, $L_b$. This is further comprised of terms with two derivatives, terms with four derivatives acting on different combinations of 2, 3 and 4 factors of b.

\beq
L_b =- d \Bigg( \HLs{\frac{2}{b} \bbox{b}} \HLK{\ + \  \frac{d-1}{b^2} (\partial b)^2} \Bigg) +  L_{b2} + L_{b3} + L_{b4} \, ,
\eeq

where the two-factor term is

\beq \label{eq:dimred-Lb2}
L_{b2} = \HLyx{\frac{d(d c_2 + 4 c_3)}{b^2} (\partial_\alpha \partial^\beta b)(\partial^\alpha \partial_\beta b) + \frac{d( 4 d c_1 + c_2)}{b^2} \bbox{b}^2} \, ,
\eeq

the three-factor term is

\beq
L_{b3} = \HLs{\frac{2d(d-1)(2d c_1 + c_2)}{b^3} (\partial b)^2  \bbox{b}} \, ,
\eeq

and finally the four-factor term is

\beq \label{eq:dimred-Lb4}
L_{b4} = \HLK{d(d-1)\frac{d(d-1) c_1 + (d-1) c_2 + 2 c_3}{b^4} (\partial b)^4} \, .
\eeq

\subsection{Gauss-Bonnet and Horndeski Theory} \label{sec:GBisHorn}

A special case of our action in eq.(\ref{eq:4pdaction}) is when $c_1 = c_3 = - c_2 /4 = -\xi$, leading to the ubiquitous Gauss-Bonnet combination of quadratic curvature scalars. While in four dimensions it is a total derivative, in higher dimensions or when non-minimally coupled it plays a role in gravity. Using our results above from eqs. (\ref{eq:dimred-L4}--\ref{eq:dimred-Lb4}) with this choice of coupling constants, the full 4D action (\ref{eq:reducedaction}) can be written in the form

\beq \label{eq:GBHornAction}
S = \frac{M_{4+d}^{2+d} \mathcal{V}}{2}\int \bd^4 x \sqrt{-g} \spar{L_H - \xi b^d R^2_{\text{GB}}} \, ,
\eeq

where $R^2_{\text{GB}} = R^2 - 4 R^{\mu\nu}R_{\mu\nu} + R^{\rho\mu\sigma\nu}R_{\rho\mu\sigma\nu}$ is the 4D Gauss-Bonnet term,\footnote{For notational simplicity, we will henceforth omit the subscript in $R_4$ when referring to the 4D Ricci Scalar, as the full $(4+d)$-dimensional version is of considerably less relevance. The (constant) $d$-dimensional version will remain concretely labelled $R_d$.} which, as it is non-minimally coupled via a factor of $b^d$ coming from the reduction of the metric determinant in the integration measure, is non-negligible. The other term present is $L_H$, a Horndeski theory Lagrangian

\begin{align}
L_H & = G_4(b,X) R_4 + P(b,X) - G_3(b,X) \bbox{b} + G_{4,X} \spar{\bbox{b}^2 - (\partial_\alpha \partial^\beta b)(\partial^\alpha \partial_\beta b)} \nonumber \\ &+ G_5(b,X) G_{\alpha\beta} \partial^\alpha \partial^\beta b - \frac{1}{6} G_{5,X} \spar{\bbox{b}^3 - 3 \bbox{b}(\partial_\alpha \partial^\beta b)(\partial_\beta \partial^\alpha b) + 2 (\partial_\alpha \partial^\beta b)(\partial_\beta \partial^\gamma b)(\partial_\gamma \partial^\alpha b)} \, ,
\end{align}

where

\begin{align}
P(b,X) & = K(b,X) - V(b) \, \\
K(b,X) & = \spar{2  d (d-1)-\frac{4 \xi R_d (d-2)}{b^2}} b^{d-2} X - 4 \xi d (d-1)(d-2)(d-3) b^{d-4} X^2  \, , \\
V(b) & = 2 \Lambda_{4+d} b^d -   R_d b^{d-2} + \xi \frac{(d-2)(d-3)}{d(d-1)} R_d^2 b^{d-4} \,  , \\
G_3(b,X) & = 2  d b^{d-1} - 4 \xi R_d (d-2) b^{d-3} - 8 \xi d (d-1)(d-2) b^{d-3} X \, ,  \\
G_4(b,X) & = 1 - 2 \xi R_d b^{d-2} - 4 \xi d (d-1) b^{d-2} X \, , \\
G_5(b) & = - 8 d \xi b^{d-1} \, .
\end{align}

where the kinetic term for the scalar field $b$ is denoted with the common convention of $X = - \partial^\mu b \partial_\mu b / 2$. Note the $(4+d)$-dimensional cosmological constant $\Lambda_{4+d}$ is absorbed into this action as the prefactor for the $b^d$ term (again resulting from the metric determinant) in the scalar potential $V(b)$. 

It is well known that the coupled Gauss-Bonnet term in four dimensions is in itself also a Horndeski theory. The action (\ref{eq:GBHornAction}) is hence the sum of two Horndeski theories, and hence a Horndeski theory itself. In fact, by explicitly rewriting the Gauss-Bonnet sector in Horndeski-like form, one can obtain the action in ``pure Horndeski'' form

\beq \label{eq:PureHornAction}
S = \frac{M_{4+d}^{2+d} \mathcal{V}}{2}\int \bd^4 x \sqrt{-g} L_H \, ,
\eeq

where now the DGSZ potentials $K$ and $G_n$ obtain logarithmic corrections in $X$, taking the modified form,

\begin{align} 
P(b,X) & = K(b,X) - V(b) \, \label{eq:GBDGSZstart} \\
K(b,X) & = \spar{2  d (d-1)-\frac{4 \xi R_d (d-2)}{b^2}} b^{d-2} X - 4 \xi d (d-1)(d-2)(d-3) b^{d-4} X^2 (7 - 2\ln |X|)  \, , \\
V(b) & = 2 \Lambda_{4+d} b^d -   R_d b^{d-2} + \xi \frac{(d-2)(d-3)}{d(d-1)} R_d^2 b^{d-4} \, , \\
G_3(b,X) & = 2  d b^{d-1} - 4 \xi R_d (d-2) b^{d-3} -12 d(d-1)(d-2) \xi b^{d-3} X (3 - \ln |X|) \, , \\
G_4(b,X) & = 1 - 2 \xi R_d b^{d-2} - 4 d(d-1) \xi b^{d-2} X (3- \ln |X|) \, , \\
G_5(b,X) & = - 4 d \xi b^{d-1} (2 - \ln |X|)  \, . \label{eq:GBDGSZend}
\end{align}

We hence conclude that higher-dimensional Einstein-Gauss-Bonnet gravity, with maximally symmetric extra dimensions and a metric ansatz of the form (\ref{eq:4pdmetric}), the four dimensional effective action is simply a Horndeski-class scalar-tensor theories whose particular potentials have been computed and presented above. This is the form of the theory which is typically easiest to work with, given the abundance of literature on Horndeski theories as a standardised template of many scalar-tensor theories. It also confirms without explicit calculation that the theory has stable second-order equations of motion.

\section{Some properties of the 4D effective theory}

\subsection{Specialness of the Gauss-Bonnet case}

The results of section \ref{sec:GBisHorn} show that choosing $c_1 = c_3 = - c_2 /4 = -\xi$ or any constant multiple of this leads to a 4D effective action in the form of a Horndeski theory. It turns out this is a special result in that other choices of the $c_n$ coefficients do not lead to Horndeski theories. For example, consider the four-derivative terms in eq. (\ref{eq:dimred-Lb2}).  Typically, terms with four derivatives do not correspond to second order equations of motion, but Horndeski's theory tells us that specific combinations of such higher derivative interactions can cancel out each others' pathological elements, leading to an overall stable theory. In the Gauss-Bonnet case, the specific terms in question correspond to the following terms in the Horndeski action:

\begin{align}
L_H \supset  G_{4,X} \spar{\bbox{b}^2 - (\partial_\alpha \partial^\beta b)(\partial^\alpha \partial_\beta b)} \, .
\end{align}

This identification is possible because in eq. (\ref{eq:dimred-Lb2}), the choice $c_1 = c_3 = - c_2 /4 = -\xi$ makes the coefficients of $\bbox{b}^2$ and $(\partial_\alpha \partial^\beta b)(\partial^\alpha \partial_\beta b)$ equal up to a sign difference; in the Horndeski action these two terms must appear in this form in order to cancel out parts that do not lead to second order contributions to the equations of motion. We hence see that general choices of $c_n$ do not generally lead to Horndeski-compatible theories in 4D. In fact, upon a more comprehensive study of the correspondence between the 4D effective action and Horndeski-allowed terms, one can prove that the Gauss-Bonnet combination is the only choice of $c_1$, $c_2$ and $c_3$ that will lead to a Horndeski theory. We present the proof of this uniqueness in  \ref{app:GBonlyproof}.

It is interesting that the Gauss-Bonnet term's specialness among combinations of quadratic curvature scalars manifests in this way. Note that this means that even the higher-dimensional analogue of the well-studied Starobinsky or $R^2$ theory of modified gravity

\beq \label{eq:starobcase}
S = \frac{M_{4+d}^{2+d}}{2} \int \bd^{(4+d)} X \sqrt{-G} \rpar{ R + \alpha R^2 - 2 \Lambda_{4+d}} \, ,
\eeq

would reduce in 4D to a theory that is not Horndeski. As well as the fact that the structure of the $b$ field's Lagrangian would have greater than second order equations of motion, a residual $R^2$ term would persist in the 4D action, and could in principle be conformally transformed into a second scalar field that interacts non-trivially with $b$. It is not immediately clear that this interaction would avoid pathology in $b$'s dynamics, but as the full $(4+d)$-dimensional theory should be stable, one might hypothesise that its dimensionally reduced form is no different, though testing this explicitly is beyond the scope of the present work. Given the speciality of the Gauss-Bonnet combination, its physical relevance, and the relative simplicity of its 4D effective action compared to Starobinsky theory or another choice of $c_n$, we will henceforth focus on the special case of the action (\ref{eq:4pdaction}) given by

\beq \label{eq:exDGBaction}
S = \frac{M_{4+d}^{2+d}}{2} \int \bd^{(4+d)} X \sqrt{-G} \rpar{ R - \xi \spar{ R^2  - 4_{AB} R^{AB} + R_{ABCD} R^{ABCD}} - 2 \Lambda_{4+d}} \, ,
\eeq

\subsection{Effective 4D Planck Mass and Cosmological Constant} \label{sec:effPMCC}
We will discuss below the properties of effective potential for the field $b$ and whether $b$ can in principle stabilise at all. Assuming for now that a static solution $b = b_0$ is reached and this persists until late-time cosmology, we can evaluate the effective cosmological constant and four dimensional effective Planck mass resulting from this. Directly broadcasting the 4D effective action (\ref{eq:GBHornAction}) into a form where the prefactor of the Ricci Scalar is $M_4^2(b,X)/2$ gives us the equation

\beq
M_4^2(b,X) = G_4(b,X) M_{4+d}^{2+d} \mathcal{V} \equiv  G_4(b,X) \alpha \, .
\eeq
Here we have introduced the notation $\alpha = M_{4+d}^{2+d} \mathcal{V}$ because as we have seen already in Section 2, this combination appears very often in the four--dimensional effective theory. 
Explicitly using the form of $G_4$ given in eqs. (\ref{eq:GBDGSZstart}--\ref{eq:GBDGSZend}), and evaluating at the point ($b = b_0$, $X = 0$) we find that the observed (constant) Planck-mass today would be

\beq
M_4^2 = (1 - 2 \xi R_d b_0^{d-2})  \alpha \, .
%M_4^2 = (1 - 2 \xi R_d b_0^{d-2})  M_{4+d}^{2+d} \mathcal{V} \, .
\eeq

Enforcing that this equality holds for the measured value of $M_4$, coming from say a weak-gravity experimental determination of Newton's constant, constrains one parameter of the theory. It is helpful to rewrite the 4D effective action using this expression for $M_4^2$ such that the $G_4$ coupling takes the form

\begin{align}
G_4(b,X) & = 1 - 2 \xi R_d b^{d-2} - 4 d(d-1) \xi b^{d-2} X (3- \ln |X|) \, , \nonumber \\
& = 1 - 2 \xi R_d b_0^{d-2} - 2 \xi R_d \rpar{b^{d-2} - b_0^{d-2}} - 4 d(d-1) \xi b^{d-2} X (3- \ln |X|) \, ,  \nonumber  \\
%& = \frac{M_4^2}{M_{4+d}^{2+d} \mathcal{V}} - 2 \xi R_d \rpar{b^{d-2} - b_0^{d-2}} - 4 d(d-1) \xi b^{d-2} X (3- \ln |X|) \, , \label{eq:rephraseg4}
& = \frac{M_4^2}{\alpha} - 2 \xi R_d \rpar{b^{d-2} - b_0^{d-2}} - 4 d(d-1) \xi b^{d-2} X (3- \ln |X|) \, , \label{eq:rephraseg4}
\end{align}

where in the first step we have added and subtracted $(2 \xi R_d b_0^{d-2})$, and in the second step we have used the definition of $M_4^2$. The action now can be written in a form explicitly containing the 4D Planck Mass such as

\beq
%S = \frac{M_4^2}{2}\int \bd^4 x \sqrt{-g} R \: + \: \frac{M_{4+d}^{2+d} \mathcal{V}}{2}\int \bd^4 x \sqrt{-g} \bar{L}_H \, ,
S = \frac{M_4^2}{2}\int \bd^4 x \sqrt{-g} R \: + \: \frac{\alpha}{2}\int \bd^4 x \sqrt{-g} \bar{L}_H \, ,
\eeq

where $\bar{L}_H$ is the same Horndeski Lagrangian specified by eqs. (\ref{eq:GBDGSZstart}--\ref{eq:GBDGSZend}), except $G_4$ is replaced with $\bar{G}_4$, defined by

\begin{align} \label{eq:g4bar}
\bar{G}_4(b,X) & = G_4(b,X)  - G_4(b_0,0) \, , \\
%& = G_4(b,X)  - \frac{M_4^2}{M_{4+d}^{2+d} \mathcal{V}} \, , \\
& = G_4(b,X)  - \frac{M_4^2}{\alpha} \, , \\
& = - 2 \xi R_d \rpar{b^{d-2} - b_0^{d-2}} - 4 d(d-1) \xi b^{d-2} X (3- \ln |X|) \, ,
\end{align}

where in the second line we have used (\ref{eq:rephraseg4}).\\

Similarly, one can show that the 4D effective cosmological constant for a stabilised field is simply

\beq \label{eq:splitactionc}
\Lambda = \frac{\alpha}{2 M_4^2} V(b_0) = \frac{\alpha}{2 M_4^2}  \spar{2 \Lambda_{4+d} b_0^4 -   R_d b_0^2 + \xi  R_d^2 \frac{(d-2)(d-3)}{d(d-1)}} b_0^{d-4} \, , \\
%\Lambda = \frac{M_{4+d}^{2+d} \mathcal{V}}{2 M_4^2} V(b_0) = \frac{M_{4+d}^{2+d} \mathcal{V}}{2 M_4^2}  \spar{2 \Lambda_{4+d} b_0^4 -   R_d b_0^2 + \xi  R_d^2 \frac{(d-2)(d-3)}{d(d-1)}} b_0^{d-4} \, , \\
\eeq

and this would allow us to write the action in a form

\beq
%S = \frac{M_4^2}{2}\int \bd^4 x \sqrt{-g} \rpar{R - 2 \Lambda} \: + \: \frac{M_{4+d}^{2+d} \mathcal{V}}{2}\int \bd^4 x \sqrt{-g} \bar{L}_H \, ,
S = \frac{M_4^2}{2}\int \bd^4 x \sqrt{-g} \rpar{R - 2 \Lambda} \: + \: \frac{\alpha}{2}\int \bd^4 x \sqrt{-g} \bar{L}_H \, ,
\eeq

where now the barred Horndeski Lagrangian $\bar{L}_H$ is again specified by eqs. (\ref{eq:GBDGSZstart}--\ref{eq:GBDGSZend}), this time with both $G_4$ and $V$ replaced by barred variations given respectively by (\ref{eq:g4bar}) and

\beq
\bar{V}(b) = V(b) - V(b_0) \, .
\eeq

The action (\ref{eq:splitactionc}) then has the property that the second term containing the Horndeski Lagrangian vanishes when $b = b_0$, as all the parts not proportional to derivatives of $b$ (i.e. $X$)  have been absorbed into the definitions of the effective Planck mass and cosmological constant in the first term, and $\bar{G}_4(b_0) = \bar{V}(b_0) = 0$ by construction.

\subsection{Effective potential and static solutions}

In the limit of negligible motion of the field, the Friedmann and Klein-Gordon equations (whose full forms can be found e.g. in \cite{Kobayashi:2011nu}) can be approximated at zeroth order in $\dot{b}$, and take the form

\begin{align}
6 H^2 G_4 & = V \, \\
V_{,b} + 12 H^2 G_{4,b}  & = 0 \, .
\end{align}

Using the equations of motion in this form allows us to search for static solutions of the system - those in which the field is frozen in place. These are of particular interest in the context of cases such as ours where the field represents the size of extra dimensions, as it implies their stabilisation, as well as the context of cosmology where static scalar fields in a potential behave as fluids with equation of state of $-1$. Particularly when thinking about stabilisation of the extra dimensions, we are also interested in static solutions where $b = b_0 \neq 0$ as this would imply extra dimensions of zero size, which has obvious difficulties in interpretation.

The presence of terms other than $V_{,b}$ in the Klein-Gordon equation implies that even in the static limit, the points at which the field can sustain a constant value will not be the minima of the bare potential $V$ but instead some effective potential due to a combination of $V$ and $G_4$. As we have neglected motion of the field in this, this calculation also makes no claims regarding whether the system can dynamically reach such a stable point \footnote{In section \ref{sec:Matter} we will argue that one should include the effects of matter in any such discussion of dynamics, and that the arbitrariness of how matter is coupled means that a more comprehensive future study of this needs to be done to fully understand such questions.}, simply whether stable points exist. What we can say from such an analysis is whether such static solutions exist in the first place, and if so what the sufficient, necessary or preferable conditions for their existence are in terms of say the number of dimensions and the combination of model parameters such as $\xi$ and $R_d$ \footnote{Narrowing down the parameter space now to only those models in which an interesting static solution exists will hence be an important building block towards the aforementioned future study of matter-inclusive dynamics.}. In this limit of negligible field motion, the static solutions we find will also be stable to zeroth order in $\dot{b}$ as long as they occur at minima of the effective potential (e.g. in the slice $\dot{b} = 0$. Their stability to small perturbations in the field velocity, however, is also dependent on the matter coupling, and can't be fully addressed within the scope of the present work.

In conventional scalar-tensor theory, the equation of motion is just $V_{,b} = 0$, and so it is natural to define an effective potential gradient $V_{\text{eff} , b}  = V_{,b} + 12 H^2 G_{4,b}$ to mimic this form. Using the Friedmann equation to eliminate $H^2$ from this expression and using the specific forms of $V$ and $G_4$ for our theory gives the effective potential gradient as

\begin{align}
V_{\text{eff} , b} & = V_{,b} + 2  \frac{G_{4,b}}{G_4} V \, ,  \nonumber \\
& = 2 d \Lambda_{4+d} b^{d-1} -   (d-2) R_d b^{d-3} + \frac{(d-2)(d-3)(d-4)}{d(d-1)}\xi  R_d^2 b^{d-5} \nonumber \\
& - \frac{4(d-2)\xi R_d}{1 - 2 \xi R_d b^{d-2}} \spar{2 \Lambda_{4+d} b^{2d-3} -   R_d b^{2d - 5} + \frac{(d-2)(d-3)}{d(d-1)} \xi  R_d^2 b^{2d -7}} \label{eq:veffbint} \\
& = \frac{1}{1 - 2 \xi R_d b^{d-2}} \Bigg[ 2 d \Lambda_{4+d} b^{d-1} -   (d-2) R_d b^{d-3} + \frac{(d-2)(d-3)(d-4)}{d(d-1)} \xi  R_d^2 b^{d-5} \nonumber \\
& - 4(3d-4) \Lambda_{4+d}\xi R_d b^{2d-3} + 6(d-2)  \xi R_d^2 b^{2d - 5} - 2 \frac{(d-2)(d-3)(3d-8)}{d(d-1)} \xi^2 R_d^3 b^{2d -7} \Bigg] \, . \label{eq:veffb}
\end{align}

Finding solutions to $V_{\text{eff} , b} = 0$ to identify the existence of static solutions, using the effective potential as given by (\ref{eq:veffb}), then reduces to a problem of solving a polynomial equation of six terms with orders $(d-1)$, $(d-3)$, $(d-5)$, $(2d-3)$, $(2d-5)$, and $(2d-7)$ . This is, of course, not generally possible for either general $d$ nor for any specified value of $d$. Below we will look at each choice of $d$ as a special case and categorise the possible behaviours.

Before proceeding to this, however, we can also directly integrate the expression (\ref{eq:veffbint}) or (\ref{eq:veffb}) with respect to $b$ to see what the effective potential $V_\text{eff}$ itself looks like, i.e. what corrections do the presence of a $G_4$ term impose on the shape of the potential felt by the field in the limit of negligible motion. This procedure yields the expression,

\begin{align}
V_\text{eff} & = 2 \Lambda_{d+4} b^d -  3 R_d b^{d-2} + \xi \frac{(d-2)(d-3)}{d(d-1)} R_d^2 b^{d-4} - \frac{1}{\xi} \log\rpar{1 - 2 \xi R_d b^{d-2}}\, \nonumber \\
& - 2 \xi^2 R_d^3 \rpar{\frac{(d-2)^2}{d(d-1)}} \times  {}_2F_1\rpar{1,\frac{2(d-3)}{d-2};\frac{3d-8}{d-2};2 \xi R_d b^{d-2}} b^{2(d-3)}  \, \nonumber \\
& - 4  \xi  R_d \Lambda_{d+4}\rpar{\frac{d-2}{d-1}} \times {}_2F_1\rpar{1,\frac{2(d-1)}{d-2};\frac{3d-4}{d-2};2 \xi R_d b^{d-2}} b^{2(d-1)}  \, , \label{eq:Vefffull}
\end{align}

where ${}_2F_1$ is the hypergeometric function. Note the structure of this potential; the first three terms are as in the bare potential, with the difference that the coefficient of the second term is now 3 instead of 1. Furthermore, logarithmic and hypergeometric correction terms are generated. The logarithmic term in particular imposes a boundary on the values $b$ may take if the reality of the potential is enforced, that is 

\beq
2 \xi R_d b^{d-2} < 1 \, .
\eeq

This is trivially satisfied if the sign of the combination of terms on the left hand side is negative, but when it is positive there will generally be some limit on the value of $b$. As well as representing a divergence in the effective potential, this inequality also imposes essentially the positivity of $G_4$ and hence that of $H^2$ in the Friedman equation (assuming $V$ is also positive) as well as that of the effective square Planck mass, a topic that will be more explicitly discussed in section \ref{sec:effPMCC}.

We would also like the stabilised system to admit a positive effective cosmological constant in four dimensions. Inspecting the Friedman equation or the effective action, one can see that the ratio of the signs of $V$ and $G_4$ will determine the sign of the effective cosmological constant, and we will again see this more explicitly in section \ref{sec:effPMCC}. As we previously discussed, $G_4$ is constrained to be positive by enforcing positivity of the square Planck mass, and so $V$ must also be positive at the point $b = b_0$ of the static solution for it to satisfy this. For now we will simply search for criteria the parameters must fulfil to permit a positive effective cosmological constant, rather than find bounds on how small the parameters must be to actually replicate the observed cosmological constant as our foremost interest here is to exclude models and numbers of dimensions on grounds of wider criteria rather than precise and comprehensive parameter searches.

\subsubsection{Case of $d$ = 1 or 2}

In $d = 1$ the expressions derived in this work are largely simplified by the fact that a one-dimensional space has zero Ricci curvature. One result of this is that the effective potential is just the bare potential, which in the limit of $R_d = 0$ is simply $V = 2 \Lambda_5 b$. Similarly, the expression for the effective potential in eq. (\ref{eq:Vefffull}) is invalid for the case $d = 2$, as in this case $G_{4,b} = 0$ and the hence the effective potential is identical to the bare potential, which takes the form $V = 2 \Lambda_6 b^2 - R_d$. Neither of these cases of are of interest to us, as the $d=1$ case has a linear effective potential which cannot stabilise, and the $d=2$ case has a quadratic potential which is capable of stabilising but only at $b = 0$. We hence rule out these models.

\subsubsection{Case of $d \geq 6$}

Inspecting the expression for $V_{\text{eff},b}$ in eq. (\ref{eq:veffb}), we see that for $d \geq 6$, $b = 0$ will always be a solution of the Klein-Gordon equation, implying a turning point of the potential necessarily exists at $b = 0$. This alone would not be a problem, if not for the fact that in models with $d \geq 6$ one also finds that $V_\text{eff}(0) = 0$ and that the effective cosmological constant at the point $b = 0$ is also zero (as this is set by the bare potential). These two facts combined mean that if there exists a stable potential minimum point other than $b=0$, then it is either not a global minimum, or it possesses a negative cosmological constant. This does not strictly rule out the case of $d \geq 6$ for interesting applications; it is entirely possible that we have a metastable state represented by a non-global minimum in effective vacuum energy. However, as this is an additional complication, we are less inclined to explore this possibility in our initial survey of models. When we, in a future work, return to the question of dynamical stabilisation, such metastable points will be undesirable as we would not expect them to be reached from generic initial conditions.

An additional complication with the class of $d \geq 6$ models is that the leading term in the $V_{\text{eff},b}$ polynomial is of order $(2d-3)$, that is, even for the case $d = 6$ we would have to deal with a potential with up to nine possible turning points whose $b$ values can not in general be analytically determined. 

Given these observations, while we do not absolutely rule out models of $d \geq 6$, we initially choose to relegate them as more convoluted and complex than strictly necessary. We will see next that the remaining cases of $d$ between 3 and 5 provide enough feasible behaviour that we need not delve into this possibility at this stage.

\subsubsection{Case of $d$=3}

In $d=3$, the integer arguments of the hypergeometic functions in the effective potential allow us to write it in the simplified form

\beq
\frac{2}{\alpha} V_\text{eff}^{(d=3)} = -\frac{\xi^2 R_d^3}{3} + \rpar{\frac{\Lambda_7}{\xi^2 R_d^2} - 3 R_d} b + \frac{\Lambda_7}{\xi R_d} b^2 + \frac{20 \Lambda_7}{6} b^3 + \rpar{\frac{\Lambda_7}{2 \xi^3 R_d^3} - \frac{1}{\xi}} \log\rpar{1 - 2 \xi R_d b} \, ,
\eeq

in which the hypergeometric functions can now be seen to simply change the prefactor of the logarithmic correction to the polynomial bare potential. Manipulation of eq. (\ref{eq:veffb}) additionally reveals that the turning points of the potential occur at $b$ values that are solutions of the cubic equation

\beq
b^3 - \frac{3}{10 \xi R_d} b^2 - \frac{3 R_d}{10 \Lambda_7} b + \frac{1}{20 \xi \Lambda_7} = 0 \, .
\eeq

As the logarithm in the effective potential imposes $2 \xi R_d b < 1$, $G_4$ will always be positive for viable field values, and so the sign of the effective cosmological constant is determined entirely by the bare potential, which in this case looks like

\beq
V^{d=3}(b) = \rpar{2 \Lambda_7 b^2 -  R_d} b\, . \label{eq:barepot3}
\eeq

As there are an odd number of extra dimensions in this case, it works out that the potential and effective potential take these asymmetrical forms. Considering we wish to interpret $b$ as a scale factor of the extra dimensions, common sense suggests we should impose $b>0$ to facilitate this. \footnote{As the metric determinant in the full action brings a factor of $b^d$ into the action, negative scale factors would imply that in odd numbers of dimensions the action would pick up a difference in sign relative to even dimensionality or positive $b$.} We therefore ignore the possibility of negative $b$. Whether the time-evolution of the system, taking matter couplings into account, obeys this is of course left to future work probing the dynamics of our model.

The simplest way to ensure a positive effective cosmological constant is possible, from eq. (\ref{eq:barepot3}), would be to impose $\Lambda_7 > 0$ and $R_d < 0$.  However, by Descartes's Rule of Signs, we would require in this case that $\xi < 0$ as a necessary and sufficient condition for a positive-$b$ turning point to exist. The discriminant of the cubic equation can also be shown to be negative for this combination of signs, identifying this as the sole real root. Then, as $R_d < 0$, and

\beq
V_{\text{eff} , b}(0) = - R_d \, ,
\eeq

is hence positive, that single turning point at positive-$b$ will always be a maximum rather than a minimum, making this case of no use. If we instead choose $R_d > 0$ to ensure that the first turning point at positive $b$ (should it exist at all) be a minimum, then $\Lambda_7 > 0$ is also a necessary but not sufficient condition for the positivity of $V$ at positive $b$ values. That is, from (\ref{eq:barepot3}) we see that $V$ will only be positive for $b^2 > R_d / \Lambda_7$. Finally, from arguments again based on Descartes's Rule, one can only guarantee the existence of a positive turning point in this situation when $\xi < 0$, and that there will only be one turning point. The gradient of the effective potential at the point $b^2 = R_d / \Lambda_7$ is 

\beq
V_{\text{eff} , b}(V = 0) =  R_d \rpar{5 - \frac{4 b \xi R_d}{1 - 2 b \xi R_d}}\, ,
\eeq

which, for $\xi < 0$ is always positive. This implies that, as there has been a change in sign of the effective potential's gradient between $b = 0$ and $b^2 = R_d / \Lambda_7$, that the minimum occurs at a $b$ value in this interval where by (\ref{eq:barepot3}) the effective cosmological constant will be negative. Conversely, if one imposes $\xi > 0$ then there will be either zero or two positive-$b$ turning points. In the case where two exist (excluding the 0-case as uninteresting), the first of these will still be a minimum but will fail to have positive cosmological constant for the same reason as the $\xi < 0$ case, and the second turning point will be a maximum. 

To conclude, the $d=3$ case looks initially promising compared to the other possible numbers of dimensions we have looked at above, but upon a closer analysis, is unable to sustain both a positive-$b$ minimum in the effective potential and a positive-valued effective cosmological constant at that minimum, making it unsuitable to produce a realistic model of our universe with stabilised extra dimensions.

\subsubsection{Case of $d$=4}

Similarly to the previous case, the hypergeometric functions have integer arguments in $d=4$, and hence the potential can once again be simplified to the form

\beq \label{effpot4}
\frac{2}{\alpha} V_\text{eff}^{(d=4)} = \frac{\xi R_d^2}{6} + \rpar{\frac{2 \Lambda_8}{\xi R_d} - 3 R_d} b^2 + 4 \Lambda_8 b^4+ \rpar{\frac{\xi R_d^2}{3} - \frac{1}{\xi} + \frac{\Lambda_8}{\xi^2 R_d^2}} \log\rpar{1 - 2 \xi R_d b^2} \, .
\eeq

As expected for an even-$d$ case, the potential only depends on powers of $b^2$ and is hence symmetric under $b \rightarrow -b$, and we do not need to worry about the physical meaning of the sign of $b$. We can, as before, also find a polynomial from (\ref{eq:veffb}) whose solutions are the turning points of this potential, and find that in this case the polynomial in question is quintic. Despite this, however, it is mathematically simpler than the $d=3$ case as one can see that $b = 0$ is always a solution, and the remaining four solutions are given by the quartic equation 

\beq
b^4 - \rpar{\frac{3 R_d}{8 \Lambda_8} + \frac{1}{4 \xi R_d}} b^2 + \frac{3 + 2 \xi^2 R_d^2}{48 \xi \Lambda_8} = 0 \, ,
\eeq

which is really just a quadratic in $b^2$, rendering this simpler to analyse than even the $d=3$ case. Casting this in the form $(b^2 - r_1)(b^2 - r_2) = 0$, the four solutions are $\pm \sqrt{r_1}$ and $\pm \sqrt{r_2}$. The signs of $r_1$ and $r_2$ thus determine if there are 0, 2 or 4 turning points apart from the guaranteed one at $b=0$. By comparison of coefficients we find the relations

\begin{align}
r_1 r_2 & = \frac{3 + 2 \xi^2 R_d^2}{48 \xi \Lambda_8} \, , \\
r_1 + r_2 & =  \rpar{\frac{3 R_d}{8 \Lambda_8} + \frac{1}{4 \xi R_d}} = \frac{R_d}{\Lambda_8}  \rpar{\frac{3}{8} + \frac{\Lambda_8}{4 \xi R_d^2}}  \, ,
\end{align}

which suggests that there are two extrema if $\xi$ and $\Lambda_8$ have different signs (regardless of $R_d$), and none if $\xi$ and $\Lambda_8$ are of the same sign but $R_d$ is different. In the case where all three parameters share a sign, there may either be four or zero extrema depending on the relative size of the parameters

Returning to the turning point at $b = 0$, the second derivative of the effective potential at this point,

\beq
V_{\text{eff},bb}^{d=4}(0) = - \alpha R_d\rpar{\frac{2}{3} \xi^2 R_d^2 + 1} \, ,
\eeq

suggests that it will be a maximum if $R_d$ is positive, and a minimum if $R_d$ is negative. The former of these is preferable as a $b=0$ minimum is phenomenologically undesirable, and it further guarantees that if any other extrema exist, at least the one closest to 0 will be a minimum instead. It is then sufficient to confirm the existence of other extrema as a condition for there to be a minimum. A positive Ricci curvature of the extra dimensions in also desirable in that the volume of the extra dimensions can then be thought of as a finite volume of a hypersphere.

\begin{figure}[t]
    \centering
    \includegraphics[width=\textwidth]{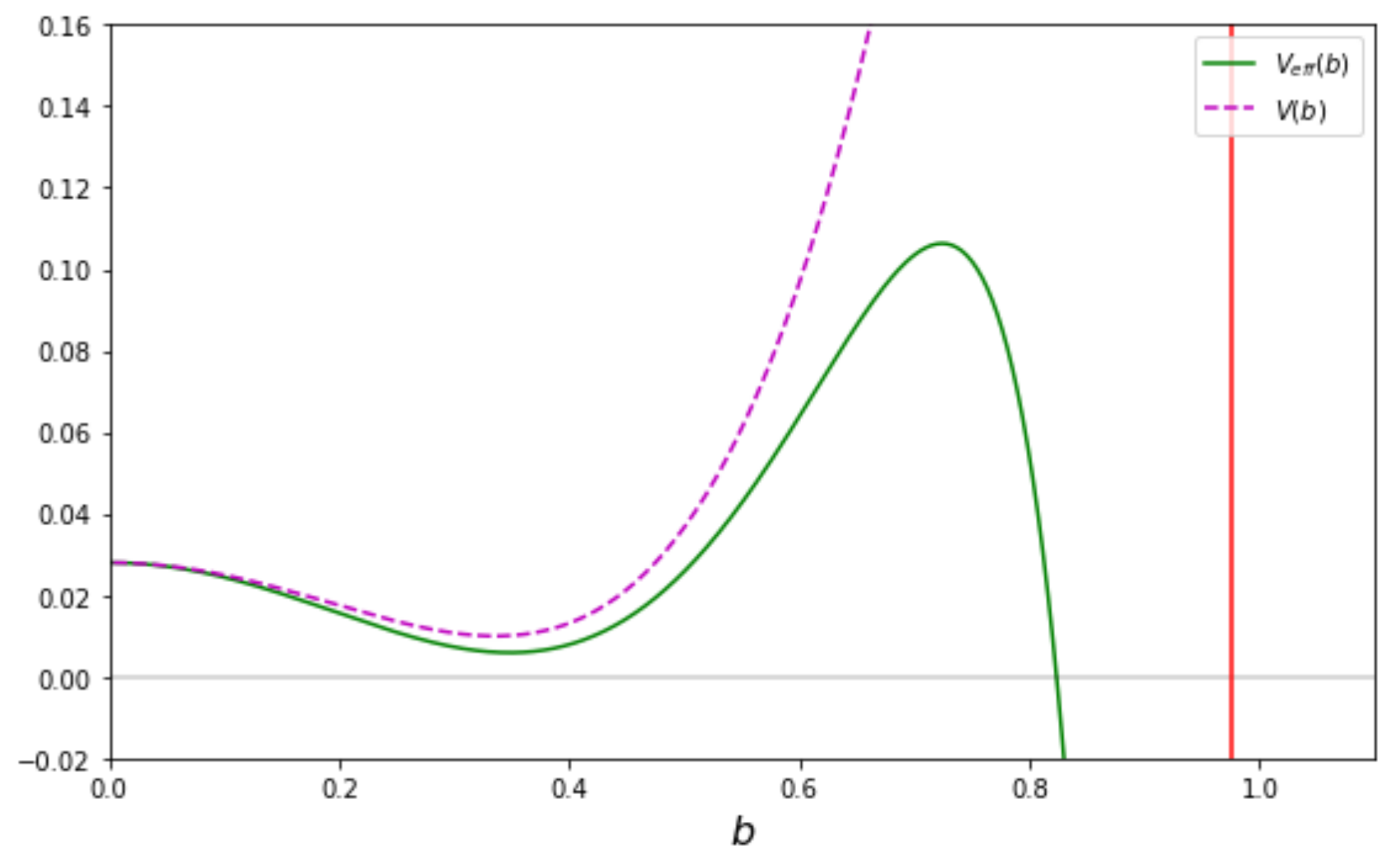}
        \caption{The effective potential (solid line) and bare potential (dashed line) for the parameter choice $\Lambda_{8} = 0.71$, $R_d = 0.32$ and $\xi = 1.64$ with $d=4$. The red line on the right shows the limiting values for $b$ for which the argument in the logarithmic term in the expression for the effective potential (\ref{effpot4}) is positive. Hence, in all realistic situations, the value of $b$ will vary between these extreme values.}
    \label{fig:Fig1}
\end{figure}

Assuming positive $R_d$, existence of other turning points would require that at least one of $\xi$ or $\Lambda_8$ are positive by the above analysis. Another condition comes from the form of the bare potential (which, as $G_4$ is positive for allowed field values, determines the sign of the effective cosmological constant),

\begin{align}\label{barepot4}
V^{d=4}(b_0)  = 2 \Lambda_8 b_0^4 - R_d b_0^2 + \frac{\xi R_d^2}{6} \, 
 = 2 \Lambda_8 \rpar{b_0^2 - \frac{R_d}{4 \Lambda_8}}^2 + \frac{R_d^2}{2} \rpar{\frac{\xi}{3} - \frac{1}{4 \Lambda_8}}
\end{align}

which we desire to be positive. We can see that this will not generally be true but for particular parameter choices, it may be. Numerical investigation reveals that the most promising scenario is when both $\xi$ and $\Lambda_8$ are positive, as it is under these conditions that one can most readily construct models where a positive cosmological constant minimum is present. Figure \ref{fig:Fig1} shows a specific example of this with parameters $\Lambda_{8} = 0.71$, $R_d = 0.32$ and $\xi = 1.64$. The dashed (purple) line shows the potential as it appears in the Friedmann equation (\ref{barepot4}), whereas the solid (green) line shows the effective potential (\ref{effpot4}). Note that we consider $b>0$ (as the scale factor is positive) and that the effective potential is not bounded from below. Dynamically we expect $b$ to grow from small values and settle down at the minimum of the effective potential. We believe that the effective potential, derived under the assumption that the time--derivatives of the field $b$ are negligible, is a good approximation for the case that the field $b$ is slowly varying, but a detailed study of the full dynamics of $b$ is beyond the scope of the paper. 

\subsubsection{Case of $d$=5}

In contrast to the previous cases of $d=3$ and $d=4$, no such simple form for the effective potential exists for $d=5$ as the arguments of the hypergeometric function are now non-integer. The $d=5$ case also has turning points corresponding to the roots of a seventh-order polynomial, which hence cannot be found in general. While we have no physical reasons to particularly neglect the case of $d=5$, it also offers no apparent advantages over $d=3$ and $d=4$ which permit much more minimalistic and simple analysis. Qualitatively, we might also expect it to behave more like $d=3$ than $d=4$ on parity grounds. While we hence note here that numerical studies of $d=5$ might unveil useful behaviour, we do not undertake that work here.

\subsection{Inclusion of matter} \label{sec:Matter}
So far we have considered a higher dimensional theory of gravity containing quadratic curvature scalars which reduces to a scalar-tensor theory in 4D. Effects resulting from the presence of matter fields in addition to this have been neglected, and while they are beyond the main scope of this work, we will briefly consider them here. First, by directly adding a minimally-coupled (for simplicity) matter term dependent on both coordinates $x$ and $y$ to eq. (\ref{eq:4pdaction}), 

\beq 
S = \frac{M_{4+d}^{2+d}}{2} \int \bd^{(4+d)} X \sqrt{-G} \rpar{ R + c_1 R^2 + c_2 R_{AB} R^{AB} + c_3 R_{ABCD} R^{ABCD} - 2 \Lambda_{4+d} + L_\text{m}(x,y)} \, ,
\eeq

we immediately come across a problem in the dimensional reduction procedure in that the integrating-out of $y$-dependent terms can no longer proceed fully. Instead we find

\beq
S = \frac{\alpha}{2} \int \bd^4 x \sqrt{-g} b^d L_{4\text{D}} + \frac{M_{4+d}^{2+d}}{2} \int \bd^4 x \ \bd^d y \sqrt{-g(x)} \sqrt{\gamma(y)} b(x)^d L_\text{m}(x,y) ,\, ,
\eeq

such that the mixed $x$ and $y$ dependencies of the matter Lagrangian prevent further simplification in the absence of an ansatz further specifying the nature of the matter such as $L_\text{m} = L_\chi(x)\delta(y-y_0)$ where $\delta$ is the Dirac delta function. Proceeding with this ansatz, the procedure of dimensional reduction and identification with a Horndeski theory as in section \ref{sec:GBisHorn} would then yield an action

\beq \label{eq:4Deffwmatter}
S = \frac{\alpha}{2}\int \bd^4 x \sqrt{-g}\spar{ L_H + C_0 b^d L_\chi} \, ,
\eeq

where $C_0$ is a constant given by

\beq
C_ 0 = \frac{\int \bd^d y \sqrt{\gamma} \ \delta(y-y_0)}{\int \bd^d y \sqrt{\gamma}} \, .
\eeq

The 4D effective action then contains $x$-dependent matter in the form of $\chi$ fields with a non-minimal coupling to the Horndeski field $b$. This implies that our scalar-tensor theory is written in the ``Einstein'' Frame where the energy momentum tensor associated with the $\chi$ fields is not covariantly conserved due to explicit interaction with the scalar. That is, variation of the above action gives

\begin{align}
T_{\mu\nu} & = T_{\mu\nu}^{(b)} + T_{\mu\nu}^{(\bar{\chi})} \nonumber \\ 
& = T_{\mu\nu}^{(b)} + C_ 0 b^d  T_{\mu\nu}^{(\chi)} \, ,
\end{align}

where $T_{\mu\nu}^{(\bar{\chi})}  = C_0 b^d  T_{\mu\nu}^{(\chi)}$ is a re-defined energy-momentum tensor which, by total energy-momentum conservation, must obey

\beq
\nabla^\mu T_{\mu\nu}^{(b)}  = - \nabla^\mu T_{\mu\nu}^{(\bar{\chi})} = Q_\nu \, .
\eeq

The bare energy-momentum tensor for $\chi$ however is independent of $b$ and only interacts with it via rescaling by a factor of $C_0 b^d$. We would hence expect that

\beq
\nabla^\mu T_{\mu\nu}^{(\chi)} = 0 \, .
\eeq

Combining this with the explicit covariant differentiation of $T_{\mu\nu}^{(\bar{\chi})}$:

\beq
\nabla^\mu T_{\mu\nu}^{(\bar{\chi})} = C_ 0 b^d \rpar{\nabla^\mu T_{\mu\nu}^{(\chi)}} + d C_0 b^{d-1} \rpar{\partial^\mu b} T_{\mu\nu}^{(\chi)} \, ,
\eeq

we would conclude that $Q_\nu = d C_0 b^{d-1} b^{,\mu} T_{\mu\nu}^{(\chi)}$. The equation of motion for $b$ then take the form

\beq
\nabla^\mu T_{\mu\nu}^{(b)} = d C_0 b^{d-1} b^{,\mu} T_{\mu\nu}^{(\chi)} \, .
\eeq

In a cosmological background, where the left-hand side produces the usual Klein-Gordon equation terms (whose particular forms are well documented for Horndeski theories), this means that the right-hand side of the equation of motion for $b$ would be not zero, but instead

\beq \label{eq:mattercoupledKGE}
\dot{\rho}_b + 3 H (\rho_b + p_b)  = -d C_0 b^{d-1} \dot{b} \rho_\chi \, ,
\eeq

while the energy density of the matter field would obey the usual fluid equation

\beq
\dot{\rho}_\chi + 3 H( \rho_\chi + p_\chi) = 0 \, ,
\eeq

and the rescaled matter energy density corresponding to the energy-momentum tensor $T_{\mu\nu}^{(\bar{\chi})}$ via covariant conservation would obey

\beq
\dot{\rho}_{\bar{\chi}} + 3 H (\rho_{\bar{\chi}}  + p_{\bar{\chi}} )  = d C_0 b^{d-1} \dot{b} \rho_\chi  = d \frac{\dot{b}}{b} \rho_{\bar{\chi}}  \, ,
\eeq

where the second equality uses $T_{\mu\nu}^{(\bar{\chi})} = C_0 b^d T_{\mu\nu}^{(\chi)}$ to rewrite the interaction term also in terms of the $\bar{\chi}$ picture. What we see from these calculations is that the presence of matter would in general affect the dynamics of the extra-dimensional scale factor in this theory. As this effect is dependent on the model of the matter fields, including the possibility of more complicated effects arising from non-minimal coupling in the $(4+d)$-dimensional action, we will neglect its presence in the name of simplicity (the bare field theory already contains 6 free parameters) and model independence (not having results depend on a specific realisation of matter) but at the cost of generality. This means our results will only strictly be applicable to situations in which the matter term in eq. (\ref{eq:mattercoupledKGE}) is negligible. The inclusion of matter will, however, influence the location of any effective potential minima as well as potentially alter the dynamics of the system which determine whether such minima are in practice reached from physically useful initial conditions or not. We leave an in-depth look at these possibilities for the future, emphasising that the purpose of the present work is to assess the initial feasibility of this theory first in the simplest available context and in turn provide a starting point for a future work which does systematically consider matter interactions.

A further complication to inclusion of matter which is not as immediate as the dynamical effects, particularly if we were to consider a non-minimal matter coupling in the $(4+d)$-dimensional action, is that one would have to take into account the point that the conformal rescaling of the metric $G$ would alter the meaning of the function $b$. In particular, a constant $b$ in one conformal frame does not necessarily mean a constant $b$ in other frames, and one would have to be careful when defining things like ``static'' extra dimensions to specify in which frame they are static and what the physical meaning of this is. On the other hand, if one were to conformally transform a 4D effective action such as eq. (\ref{eq:4Deffwmatter}) to remove the matter-scalar coupling, one would only have to conformally rescale the metric $g$ to achieve this, not the full metric $G$,  leaving $b(t)$ unaltered from it's original definition, so it may be possible to avoid such complications.

\section{Conclusion}

In this paper we considered a gravitational action in $(4+d)$ dimensions containing the usual Einstein-Hilbert and cosmological constant terms as well as the squares of the Ricci scalar and the Ricci and Riemann tensors with arbitrary constant coefficients. By invoking an ansatz that the metric is the product of a normal four-dimensional metric $g(x)$ and a maximally-symmetric $d$-dimensional metric $\gamma(y)$ possessing a scale factor $b(x)$, we obtained the 4D effective theory by dimensional reduction. The resulting 4D effective action is a complicated scalar-tensor theory with coupling constants and potential terms dependent on the extra-dimensional Ricci curvature and volume, as well as the parameters of the full theory. 

A special case of our results is found when the coefficients of the quadratic curvature scalars in the full action are in the famous Gauss-Bonnet ratio. This combination uniquely leads to a 4D effective action which is a Horndeski theory, albeit with highly non-trivial potentials. A corollary of this is that even arguably-simpler theories like Starobinsky $R^2$ gravity in $(4+d)$ dimensions do not correspond to Horndeski-class scalar tensor theories in 4D. As the most interesting and simple possibility, we proceeded to focus on this special case.

In the absence of matter fields, we asked the question of whether the effective potential of the field $b$ possesses a minimum. That is, does there exist a point at which the extra dimensions could stabilise? We further impose the requirement that at such a stable point, if it may exist, that the effective 4D cosmological constant should be positive-definite. The feasibility of this depends on the number of extra dimensions, and the signs and values of the bare cosmological constant, the Gauss-Bonnet coupling and the extra-dimensional curvature. The cases $d=4$ and $d=5$ are the most promising, with $d=4$ in particular being mathematically more easily handled, and so we focus particularly on this example here, showing that with positive extra-dimensional curvature and appropriately sizes of other parameters, one can find examples where a positive-$\Lambda$ minimum exists. Meanwhile $d=1$ and $d=2$ are mostly trivial and unable to reliably stabilise, $d=3$ can only sustain negative-$\Lambda$ solutions, and $d \geq 6$ may be able to produce desirable solutions with the caveat that they will be unavoidably metastable at best.

We then briefly discussed the issue of including a matter term in the full theory. We studied briefly a toy model of matter which possesses only delta function dependence on $y$, thus isolating the matter to a single 4D hypersurface, and proceeded to derive the fluid equations for the scalar field $b$ and the matter density in this scenario. While a more realistic treatment of matter fields are needed and should be pursued in future work, it is clear that the details of any matter coupling will affect our results, but there is too much freedom in how this is done to conduct a comprehensive investigation of the possibilities at this stage. 

While we have hence investigated the question of whether stable points can even exist in this theory, we have not answered the accompanying question of whether such a stable point is dynamically approached or not from realistic or generic initial conditions. A future dynamical system analysis, with a reasonable treatment of matter, is left to future work (see \cite{Canfora:2013xsa, Canfora:2014iga, Pavluchenko:2017svq} for work in this direction). It would be also important to study whether inflation, driven by the field $b$, is a viable option. Here, however, our initial results provide a promising starting point, motivating this future work and justifying the time and effort that would need to be spent on this endeavour, given the complexity of the system. 

\vspace{0.5cm}

{\bf Acknowledgements:} CvdB is supported (in part) by the Lancaster-Manchester-Sheffield Consortium for Fundamental Physics under STFC grant: ST/L000520/1. CL was supported by a STFC studentship.

\pagebreak

\appendix

\section{Proof that the Gauss-Bonnet combination is the only choice of coefficients leading to a Horndeski theory} \label{app:GBonlyproof}

Many terms in the 4D effective action such as those proportional to $\bbox{b}$ are always Horndeski-allowed, but other terms share a common prefactor, or must have a prefactor that is a derivative of another term's in order for higher derivative parts to cancel out in the equations of motion. Particularly, this pertains to the $G_4$ section of the Horndeski action

\begin{align}
L_H \supset G_4(b,X) R + G_{4,X} \spar{\bbox{b}^2 - (\partial_\alpha \partial^\beta b)(\partial^\alpha \partial_\beta b)} \, .
\end{align}

We equate this with the pertinent terms in 4D effective action derived in section \ref{sec:model} to find conditions on the $c_n$ coefficients such that Horndeski form is achieved. Without loss of generality, we take $c_2 = A c_1$ and $c_3 = B c_1$. If the result is to be a Horndeski theory, then there should be a solution for $A$ and $B$. Proceeding, we conclude by equating the $R$-proportional part of (\ref{eq:dimred-L4}) to that of Horndeski's action that 

\beq \label{eq:dimred-L4app}
G_4 =1 + 2 c_1 \frac{R_d}{b^2} + 4 c_1 \frac{d(d-1)}{b^2} X \, ,
\eeq

and by equating the terms in (\ref{eq:dimred-Lb2}) with the $G_{4X}$ term in Horndeski's theory, we find the two equalities

\beq \label{eq:dimred-Lb2app}
G_{4X} =  \frac{d( 4 d c_1 + c_2)}{b^2} = -\frac{d(d c_2 + 4 c_3)}{b^2} \, .
\eeq

Differentiating (\ref{eq:dimred-L4app}) with respect to $X$ and equating it to the first expression in (\ref{eq:dimred-Lb2app}) yields

\beq
4 \frac{d(d-1)}{b^2} = \frac{d(4 d+ A)}{b^2}  \quad \rightarrow \quad A = -4 \, .
\eeq

Substituting this result for $A$  into (\ref{eq:dimred-Lb2app}) then yields

\beq 
 \frac{d( 4 d - 4)}{b^2} = -\frac{d(-4d + 4 B)}{b^2}  \quad \rightarrow \quad B = 1 \, .
\eeq

Hence $c_1 = c_3 = - c_2 /4 $ and this is the unique solution to the above system of equations, proving that only a constant multiple of the Gauss-Bonnet combination of quadratic curvature scalars will lead to a Horndeski theory in the 4D effective action.

\pagebreak

\bibliography{ExD-Bib}
\bibliographystyle{jhep}

\end{document}